%% file: main_arxiv.tex
\documentclass[letter,final,onecolumn,11pt]{IEEEtran}

\usepackage[utf8]{inputenc}
\usepackage{graphicx}
\usepackage{amssymb}
\usepackage[cmex10]{amsmath}
\usepackage{color}
\usepackage{theorem}
\usepackage{pifont}
\usepackage{nicefrac}
\usepackage{booktabs}
\usepackage{bbm}
\usepackage{cite}

\usepackage[english]{babel}
\usepackage[english = american]{csquotes}
\usepackage{mathtools}
\usepackage{mathdots}
\MakeOuterQuote{"} 
\usepackage[ruled,vlined,titlenumbered]{algorithm2e}

\definecolor{light_gray}{rgb}{0.6,0.6,0.6}
\definecolor{awgray}{rgb}{0.7,0.7,0.7}
\definecolor{awgray_dark}{rgb} {0.4,0.4,0.4}
\usepackage{pgf}
\usepackage{tikz}
\usetikzlibrary{matrix,chains,positioning,arrows,calc,decorations,plotmarks,patterns, fit,backgrounds}
\usepackage{pgfplots}
\usepgfplotslibrary{external}
\usetikzlibrary{external}                       % load externalization library
\tikzsetexternalprefix{tikzpic/}                   % selecte directory for externalized figures

\pgfplotscreateplotcyclelist{mylist}{red,blue,black,yellow,brown}

\tikzset{
	>=stealth',
	mycircle/.style={circle, draw=gray, very thick, text width=.1em, minimum height=1.5em, text centered}, 
	mycircle_small/.style={circle,draw=awgray_dark,fill = awgray_dark, inner sep=0,minimum size=.6em},       
	mycircle_small_black/.style={circle,draw=black,fill = black, inner sep=0,minimum size=.6em},   
	mybox/.style={rectangle,rounded corners,draw=black, thick,text width=1em,minimum height=4em,minimum width=4em,text centered},     
	mybox_small/.style={rectangle,rounded corners,draw=black, thick,text width=1em,minimum height=2em,minimum width=2em,text centered},               
	mybox_vec/.style={rectangle,rounded corners,draw=black, thick,text width=1em,minimum height=0.7em, minimum width=4em,text centered},  
	mybox_vec_short/.style={rectangle,rounded corners,draw=black, thick,text width=1em,minimum height=0.7em, minimum width=2em,text centered},  
	mybox_block/.style={rectangle,rounded corners,draw=black, thick,text width=1em,minimum height=2em,minimum width=4.75em,text centered},                  
	mybox_symbol/.style={
		rectangle,
		rounded corners=0.1em,
		draw=black, thick,
		minimum height=0.8em,
		minimum width=0.7em,
		text centered},
	mybox_column/.style={
		rectangle,
		rounded corners=0.1em,
		draw=black, thick,
		minimum height=3.9em,
		minimum width=0.7em,
		text centered}, 
	pil/.style={->, thick, shorten <=2pt, shorten >=2pt,},
}

\input{defs.tex}

\begin{document}

\title{A Public-Key Cryptosystem from Interleaved Goppa Codes}
\author{Molka Elleuch, Antonia Wachter-Zeh, Alexander Zeh
\thanks{M. Elleuch was with the Institute for
Communications Engineering, Technical University of Munich, Germany.
A. Wachter-Zeh is with the Institute for
Communications Engineering, Technical University of Munich, Germany.
A. Zeh is with Infineon Technologies Munich, Germany.
Emails: \{molka.elleuch, antonia.wachter-zeh\}@tum.de, alexander.zeh@infineon.com
This work was supported by the Technical University of Munich-Institute
for Advanced Study, funded by the German Excellence Initiative and European
Union Seventh Framework Programme under Grant Agreement No. 291763
and the German Research Foundation (Deutsche Forschungsgemeinschaft,
DFG) unter Grant No. WA3907/1-1.}}
\maketitle

\begin{abstract}
In this paper, a code-based public-key cryptosystem based on interleaved Goppa codes is presented.
The scheme is based on encrypting several ciphertexts with the same Goppa code and adding a burst error to them.
Possible attacks are outlined and the key size of several choices of parameters is compared to those of known schemes for the same security level. For example, for security level 128 bits, we obtain a key size of 696 Kbits whereas the classical McEliece scheme based on Goppa codes using list decoding requires a key size of 1935 Kbits.
\end{abstract}

%\begin{IEEEkeywords}
\begin{center}
\textbf{Keywords}:
Code-based cryptography, Goppa codes, Interleaved codes, McEliece system
\end{center}%\end{IEEEkeywords}

\begin{center}
\textbf{Mathematics Subject Classification (2000)}: 11T71 · 94A60 · 94B05 · 94B20
\end{center}

\section{Introduction}\label{sec:introduction}
The security of a cryptographic scheme is achieved due to the intractability of certain problems through modern computers. Currently, several companies are trying to build a quantum computer with computational capabilities and power beyond classical computing limits. 
%Quantum computers use mechanical phenomena like superposition in order to perform computing. While classical computing works with bits, quantum computers use quantum bits or shortly  "qubits". In comparison to classical bits, qubits can have more values than 0/1 or false/true. Quantum computers have consequently a computational power beyond classical computing power. Therefore, the security of a cryptographic scheme can be influenced by the capabilities of the computing system. It may be even threatened or broken if there exists an algorithm which solves the cryptographic problem efficiently on a certain powerful computing system, i.e. on a quantum computer. \\
%
%In 1984, Bennett and Brassard developed the first quantum cryptographic key distribution protocol \cite{Bennett}. Since then, a lot of research was carried in the field of quantum computing. 
In 1994, Shor introduced an algorithm \cite{Shor-PolynomialTimeAlgoPrimeFactorizationQuantum} that enables factorization of large integers and finding discrete logarithms on capable quantum computers efficiently.
Shor's algorithm would therefore break currently used public-key cryptographic algorithms such as RSA~\cite{RivestShamirAdleman-PKC} and also systems based on elliptic curve cryptography. 
%Two years later, Grover developed another quantum algorithm which ensures a polynomial speed-up in
%unstructured search.%, reducing the complexity from $\mathcal{O}(N)$ to $\mathcal{O}(\sqrt{N})$. 
%These algorithms could break or weaken the problems of concurrent practical cryptosystems. 
%Moreover, quantum computers of up to 72-qubits were recently built \cite{IBM, Google}. However, these quantum computers are still not able to run relevant quantum algorithms. Nevertheless, it is only a question of time till such a quantum computer will be built. 
As a result, the \emph{National Institute of Standards and Technology} (NIST) recently initiated a standardization process of post-quantum secure cryptosystems, including code-based cryptosystems.

The McEliece code-based cryptosystem \cite{McEliece-PKC} is based on the hardness of decoding a random code.
In this scheme, Alice chooses a linear code, masks its generator matrix by multiplying it with two special random matrices and keeps all three as a private key. Her public key is the \emph{product} of these matrices, which can be used for encryption. However, decryption can only be done if the decomposition of the public key into the three matrices is known. The original system based on Goppa codes is still unbroken and has been recommended by the \emph{PQCrypto project} for long-term protection against attacks by a quantum computer \cite{Augot-RecommendationPQSystem}. 
However, the key sizes are rather large compared to classical public-key cryptosystems which has motivated a large amount of research on different code classes for the McEliece system.
Many variants based on different code classes, e.g., Reed--Solomon codes were broken~\cite{SidelnikovShestakov}. Unbroken variants with smaller key sizes than when using Goppa codes include variants of the McEliece cryptosystem (or its dual variant, the Niederreiter scheme \cite{Niederreiter-Cryptosystem}) based on MDPC codes \cite{Misoczki_2013}; and some systems based on rank-metric codes: e.g., \cite{Loidreau-GPT-ACCT2016} (originally called the \emph{Gabidulin--Paramanov--Tretjakov} (GPT) system \cite{Gabidulin1991Ideals}), the repair of~\cite{FaureLoidreau-PKC-reconstructingPPoly} which was shown in \cite{WachterzehPuchingerRenner-FLCrypto-ISIT2018}, low-rank-parity-check codes \cite{GaboritRuattaSchrek-LRPC}; and systems based on twisted codes \cite{BeelenBossertPuchingerRosenkilde_TwistedRS_2018,PuchRenWa-TwistedGPT_ACCT2018}.

The original proposal of the McEliece system \cite{McEliece-PKC} using Goppa codes is considered to be secure against structural attacks. Encryption and decryption are very efficient due to efficient encoding/decoding methods of Goppa codes.
%s based on a matrix-vector multiplication. 
However, in practice the system has the severe drawback of very large key sizes (e.g., several hundred KByte for 128 bits security level). 
%Obviously, each cryptosystem has its own assets and deficits. However, implementing post-quantum algorithms and using them instead of traditional cryptosystems would be unpractical in this stage since embedded microcontrollers have very constrained memory resources. 
The main goal of this paper is therefore to reduce 
the key size of the Goppa code-based McEliece system while not changing the structure of the code. The latter is important as Goppa codes are the only class of codes that has remained resilient against structural attacks on the McEliece cryptosystem for long time. 
Our significant key size reduction is achieved by using so-called \emph{interleaved codes}, cf.~\cite{Bleichenbacher2003Decoding,Coppersmith2003Reconstructing,BrownMinderShokrollahi_ImprovedDecAGCodes-2005,Schmidt_CollaborativeDecoding_2009}, i.e., using several parallel codewords of the same code, in combination with burst errors.

This paper is structured as follows. Section~\ref{sec:prelim} gives preliminaries on Goppa codes and the McEliece system and introduces notation. 
In Section~\ref{sec:system}, the new system based on interleaved codes is defined and in Section~\ref{sec:attacks}, possible attacks are shown. This analysis includes an attack that breaks the system if the interleaving order is too large.
The key sizes and security levels of our system are compared to known systems in Section~\ref{sec:param}.
Section~\ref{sec:concl} concludes the paper and gives an outlook.

\section{Preliminaries}\label{sec:prelim}
\subsection{Goppa Codes and Reed--Solomon Codes}
%\subsection{Definitions and Notations}
Let $p$ be a prime power and let $q = p^m$ for some $m>1$.
Let $\Fp$ and $\Fq$ denote the finite fields of order $p$ and~$q$, respectively, and let $\Fq[x]$ denote the polynomial ring over $\Fq$.

We index vectors and matrices starting from one, e.g., $\vec{a} = (a_1,a_2,\dots,a_{n}) \in \Fq^n$ denotes a vector of length $n$ with coefficients in $\Fq$.
By the triple $[n,k,d]_p$, we denote a $p$-ary code of length $n$, dimension $k$, and minimum Hamming distance $d$.

%\subsection{Goppa Codes}
Let $\mathcal{L} = (\alpha_1, \alpha_2, \dots, \alpha_n) \in \Fq^n$ consist of $n$ non-zero distinct elements (code locators) of $\Fq$ with $n \leq q$. Let $g(x)\in \Fq[x]$ be an irreducible monic polynomial of degree $r$ such that $g(\alpha_i) \neq 0, \forall i=1,\dots,n$ (sometimes called the \emph{Goppa polynomial}).
For any word $\vec{e} = (e_1,e_2,\dots,e_n) \in \Fp^n$, the corresponding \emph{Goppa syndrome polynomial} ${s}_{\vec{e}}(x) \in \Fq[x]$ is defined by
\begin{equation*}
s_{\vec{e}}(x) =  \sum_{i=1}^{n} \frac{e_i}{x-\alpha_i} \mod g(x).
\end{equation*}
\begin{definition}[Goppa Code]
	The (irreducible) \emph{Goppa code} $\Gamma(\mathcal{L},g)$ with support $\mathcal{L}$ is the kernel of the syndrome function over $\Fp$, i.e., the set:
	\begin{equation}
	\Gamma(\mathcal{L},g) \triangleq \left\{\vec{e} \in \Fp^n : s_{\vec{e}}(x) = 0 \mod g(x)\right\}.
	\end{equation}
\end{definition}
The Goppa code $\Gamma(\mathcal{L},g)$ is an $[n, k\geq n-mr, d\geq r+1]_p$ code (cf.~\cite[Ch.~12]{MacWilliamsSloane_TheTheoryOfErrorCorrecting_1988}).
For binary Goppa codes, it can be shown that their minimum distance is at least $2r+1$.

Binary Goppa codes can be decoded up to $r$ errors uniquely by using Patterson's algorithm \cite{Patterson_GoppaDecoding}. For non-binary Goppa codes, only $\lfloor\frac{r}{2}\rfloor$ errors can be guaranteed to be decoded uniquely, but the algorithm from \cite{Barreto_SquareFreeGoppa} can decode up to $\lfloor\frac{2r}{p}\rfloor$ errors with high probability which improves upon Patterson's algorithm only for $p=3$.

Note that the results of this paper extend straight-forward to square-free Goppa codes (i.e., $g(x)$ has no multiple roots), but for simplicity, we restrict the description to irreducible Goppa codes.

The Goppa code $\Gamma(\mathcal{L},g)$ is a $p$-ary subfield subcode of a generalized Reed--Solomon (RS) $\RS{n}{\kRS=n-r}$ code over $\Fq$ of length $n$, dimension $\kRS$ and minimum distance $r+1$ which is defined by:
%An RS code \RS{n}{k} of length $n$ and dimension $k$ over $\Fq$ with $n<q$ is given by
\begin{equation} \label{eq:RSsuper}
\RS{n}{\kRS} = \{ (\nu_1 f(\alpha_1),\nu_2f(\alpha_2),\dots,\nu_nf(\alpha_n)) : f(x) \in \Fq[x], \deg f(x) < \kRS \},
\end{equation}
where $\alpha_1,\dots,\alpha_n \in \Fq$ are distinct elements and $\nu_i = \frac{1}{g(\alpha_i)}$, for all $i=1,\dots,n$, cf.~\cite[p.~182]{Roth_IntroductiontoCodingTheory_2006}.
%where $\Fq [x]$ denotes the set of all polynomials with coefficients in $\Fq$. %of degree less than $k$ and indeterminate $x$.
%RS codes are known to be \emph{Maximum Distance Separable} (MDS) codes, i.e., their minimum Hamming distance is $d=n-k+1$.

\subsection{Interleaved Goppa and Reed--Solomon Codes}\label{sec:int-goppa}
In this paper, \emph{code interleaving} refers to using $s$ parallel codewords of the same code. 
An interleaved Goppa code is therefore defined as follows.
\begin{definition}[Interleaved Goppa Code] \label{def:Int-Goppa}
	Let $\Gamma(\mathcal{L},g)$ denote an $[n, k\geq n-mr, d\geq r+1]_p$ Goppa code.
    An $s$-interleaved Goppa code is denoted by $\mathcal{I}\Gamma(\mathcal{L},g,s)$ and defined by
	\begin{equation*}
	\mathcal{I}\Gamma(\mathcal{L},g,s)= 
	\left\lbrace
	\begin{pmatrix}
	\vec{c}^{(1)}\\
	\vec{c}^{(2)}\\
	\vdots\\
	\vec{c}^{(s)}\\
	\end{pmatrix}
	\right\rbrace, 
	\end{equation*}
	where $\vec{c}^{(i)}\in \Gamma(\mathcal{L},g)$, $\forall i=1,\dots,s$.
\end{definition}

Further, an \emph{interleaved (generalized) Reed--Solomon} (IRS) code can be defined as follows.
\begin{definition}[Interleaved Reed--Solomon Code] \label{def_IRSCode}
	For $n$ distinct elements $\alpha_1,\alpha_2,\dots,\alpha_n \in \Fq$ and $n$ non-zero elements $\nu_1,\nu_2,\dots,\nu_n \in \Fq$,
	an Interleaved Reed--Solomon code $\IRS{s}$ of interleaving order $s$ is given by
	\begin{equation*}
	\IRS{s}= 
	\left\lbrace
 \begin{pmatrix}
	(\nu_1 f^{(1)}(\alpha_1),\nu_2f^{(1)}(\alpha_2),\dots,\nu_nf^{(1)}(\alpha_n))\\
	(\nu_1f^{(2)}(\alpha_1),\nu_2f^{(2)}(\alpha_2),\dots,\nu_nf^{(2)}(\alpha_n))\\
	\vdots \\
	(\nu_1f^{(s)}(\alpha_1),\nu_2f^{(s)}(\alpha_2),\dots,\nu_nf^{(s)}(\alpha_n))
	\end{pmatrix}
	\right\rbrace, 
	\end{equation*}
	where $f^{(i)}(x) \in \Fq [x], \deg f^{(i)}(x) < \kRS$, $\forall i=1,\dots,s$.
\end{definition}
Notice that in general, the definition of IRS codes can be more general, i.e., having $s$ parallel codewords from $s$ RS codes with different dimensions and different column multipliers $\nu_i$. However, in this paper we focus on so-called \emph{homogeneous} IRS codes, i.e., the $s$ rows are codewords from the same RS code.
%If $k_i = k$, for all $i=1,\dots,s$, the IRS code is called \textit{homogeneous}, otherwise \textit{heterogeneous}. 

%In this contribution, 
%we frequently consider so-called \emph{burst errors}.
%This is equivalent to a transmission of the IRS code over a $Q$-ary symmetric channel, where $Q=q^s$.
%
%Let 
%\renewcommand*\arraystretch{1.3}
%{\setlength\arraycolsep{0.35em}
%	\begin{equation*}
%	\C=\left(\begin{matrix}
%	\c^{<1>}\\ \c^{<2>}\\ \vdots\\ \c^{<s>}
%	\end{matrix}\right)
%	=\left(\begin{matrix}
%	c_1^{<1>}&c_2^{<1>}&\dots&c_{n}^{<1>}&\\ 
%	c_1^{<2>}&c_2^{<2>}&\dots&c_{n}^{<2>}&\\ 
%	\vdots&&& \vdots  \\
%	c_1^{<s>}&c_2^{<s>}&\dots&c_{n}^{<s>}&\\ 
%	\end{matrix}\right)
%	\end{equation*}}
%denote the transmitted codeword of an \IRS{s} code and let
%\begin{equation*}
%\R= \left(\begin{matrix}
%\r^{<1>}\\ \r^{<2>}\\ \vdots\\ \r^{<s>}
%\end{matrix}\right)
%=\C+\E = \C + \left(\begin{matrix}
%\e^{<1>}\\ \e^{<2>}\\ \vdots\\ \e^{<s>}
%\end{matrix}\right)
%\end{equation*}
%denote the received word. 

Interleaved codes are frequently considered in connection with \emph{burst errors}, i.e.,
the $s$ elementary codewords of the $s$-interleaved code 
are affected by $s$ elementary error words $\vec{e}^{(1)},\vec{e}^{(2)},\dots, \vec{e}^{(s)}$ of weight $\wt(\vec{e}^{(i)}) = t_i \leq t$ where the union of the~$s$ sets of 
error positions $\mathcal E = \mathcal E^{(1)}\cup\mathcal E^{(2)}\cup\dots\cup\mathcal E^{(s)}\subseteq \{ 1, \dots, n\}$ has cardinality  $|\mathcal E| = t$. Equivalently, $\mathcal E$ denotes the $t$ non-zero columns of $\E$ where
\begin{equation}\label{eq:matE}
\vec{E} \triangleq 
\begin{pmatrix}
\vec{e}^{(1)}\\ \vec{e}^{(2)}\\ \vdots \\ \vec{e}^{(s)}
\end{pmatrix}.
\end{equation}
%Each elementary received word is denoted by $\mathbf{r}^{<i>} $ $= \mathbf{c}^{<i>} + \mathbf{e}^{<i>} = $
%$(r_1^{<i>},r_2^{<i>},\dots, r_n^{<i>})$, respectively each received polynomial 
%is $r^{<i>}(x)=\sum_{j=1}^n r_j^{<i>} x^{j-1}$, for all $i=1,\dots,s$. 
%
%%Assume, the codewords are affected by a burst error of weight $t$, i.e., each of the $s$ codewords is affected by an error at the same $t$ positions. 
%For a $Q$-ary symmetric channel, each error matrix $\E$ with a fixed number $t$ of non-zero columns is equi-probable.
%This error model is the same as in \cite{Bleichenbacher2003Decoding,Coppersmith2003Reconstructing,BrownMinderShokrollahi_ImprovedDecAGCodes-2005,Schmidt_CollaborativeDecoding_2009}. Note that in \cite{Coppersmith2003Reconstructing} these errors are called \emph{synchronized errors}.

By solving a joint key equation (cf.~\cite{Schmidt_CollaborativeDecoding_2009}), for an $\IRS{s}$ code,
\begin{equation*}
\tIRS \triangleq \left\lfloor \frac{s}{s+1} (n-\kRS) \right\rfloor
\end{equation*}
burst errors (i.e., erroneous columns of $\vec{E}$) can be corrected uniquely with high probability, cf.~\cite{Bleichenbacher2003Decoding,Coppersmith2003Reconstructing,BrownMinderShokrollahi_ImprovedDecAGCodes-2005,Schmidt_CollaborativeDecoding_2009}. 
Compared to unique decoding, this can increase the decoding radius by a factor of almost two.
Note that there are decoders that achieve an even higher decoding radius for small $s$ (and mostly for low-rate IRS codes), e.g., \cite{WachterzehZehBossert-InterleavedBeyondJoinError_DCC,PuchingerRosenkilde-Interleaved-ISIT2017}, 
but in this paper, we focus on decoding up to $\tIRS$ errors.
As we will see in Section~\ref{subsec:dec-guarant}, if we choose $s \geq \tIRS$, we can \emph{guarantee} decryption/decoding, but such a system is not secure (Section~\ref{sec:attacks}). In particular, for $s=n-\kRS-1$, we obtain the maximum possible decoding radius of all IRS decoders (also \cite{WachterzehZehBossert-InterleavedBeyondJoinError_DCC,PuchingerRosenkilde-Interleaved-ISIT2017}), which is $\tIRS = s = n-\kRS-1$.

If we interleave the $\RS{n}{n-r}$ code which is a supercode (see~\eqref{eq:RSsuper}) of a given $\Gamma(\mathcal{L},g)$ code with parameters $[n, k\geq n-mr, d\geq r+1]_p$, then
\begin{equation}\label{eq:tIRS}
\tIRS = \left\lfloor \frac{sr}{s+1}  \right\rfloor,
\end{equation}
and for $s = r-1$, we obtain $\tIRS = s = r-1$.
Since the $\mathcal{I}\Gamma(\mathcal{L},g,s)$ code is a subfield subcode of the $\IRSlong{n}{n-r}{s}$ code, also the $s$-interleaved Goppa code can be decoded up to $\tIRS = s = r-1$ errors.

\begin{figure*}[htb!]
	\normalsize
	\begin{minipage}[b]{.47\textwidth}
		\centering
		\includegraphics[width=0.8\linewidth]{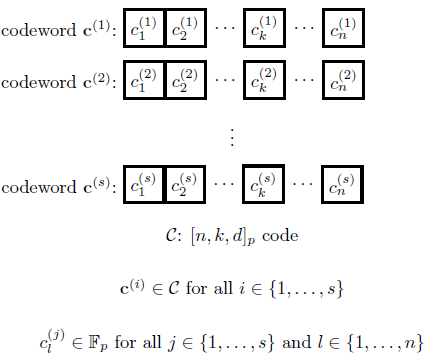}
		%\input{tikztex/decoding_regions_rank.tex}
%		\subcaption{General codes in rank metric}\label{fig:rank_decoding_region}
	\end{minipage}
	\begin{minipage}[b]{.47\textwidth}
		\centering
		\hspace{-4ex}
		\vspace{4ex}
		\includegraphics[width=1.1\linewidth]{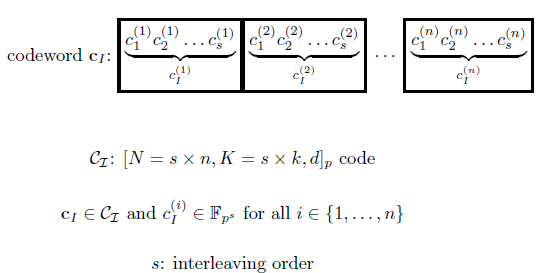}
		%\input{tikztex/decoding_regions_gabidulin.tex}
%		\subcaption{Gabidulin codes}\label{fig:gabidulin_decoding_region}
	\end{minipage}
	%\caption{Decoding regions of codes in rank metric}
	\caption{Concept of interleaving: $s$ codewords of an $[n,k,d]_p$ code are encoded in parallel (left). They can equivalently be seen as one word of length $sn$ (right side).
	\label{fig:interleaving}}
	\hrulefill
\end{figure*}

The concept of interleaving is also illustrated in Fig.~\ref{fig:interleaving}.

%\begin{figure}[htb]
%	\centering
%%	\begin{subfigure}{}
%%		\centering
%		\includegraphics[width=0.38\linewidth]{fig/IC.png}
%%		\label{fig:sub1}
%%	\end{subfigure}%
%%	\begin{subfigure}{}
%%		\centering
%		\includegraphics[width=0.53\linewidth]{fig/ICD.png}
%%		\label{fig:sub2}
%%	\end{subfigure}
%	\caption{Concept of interleaving: $s$ codewords of an $[n,k,d]_p$ code are encoded in parallel (left). They can equivalently be seen as one word of length $sn$ (right side).
%%	\textcolor{blue}{TODO: change in figure: $q$ should be $p$; indices shouldn't be bold; it should be $\mathbf{c}^{(i)}$ instead of $\mathbf{c}_i$}	
%		}
%	\label{fig:interleaving}
%\end{figure}

\newpage
\subsection{The McEliece Cryptosystem}
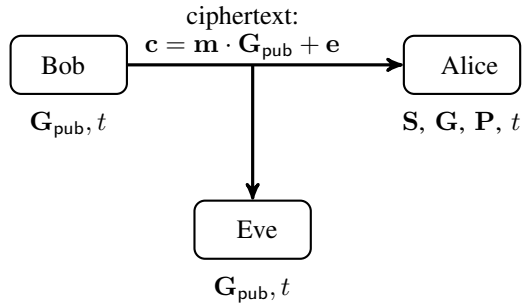
\begin{figure}[htb]
	\centering
%	\vspace{-5ex}
	\begin{tikzpicture}[scale = 1.1]
	\node[mybox_block,minimum width=4em](alice){$\hspace{-1ex}$\textcolor{black}{\small{Bob}}};
	%	\node[rectangle,left=3ex of alice](blub){};
	\node[rectangle, below=.5ex of alice](k1){$\hspace{-.1ex}$\textcolor{black}{\small{$\mathbf{G}_{\mathsf{pub}}, t$}}};	
	\node[mybox_block, right=21ex of alice,minimum width=4em](bob){$\hspace{-.5ex}$\textcolor{black}{\small{Alice}}};
	\node[rectangle, below=.5ex of bob](k2){$\hspace{-.1ex}$\textcolor{black}{\small{$\mathbf{S}$, $\mathbf{G}$, $\mathbf{P}$, $t$}}};
	\draw[->,very thick,color=black](alice.east)--(bob.west);
	\node[mybox_block, below right=8ex and 5ex of alice,minimum width=4em](eve){$\hspace{-.1ex}$\textcolor{black}{\small{Eve}}};
	\node[rectangle, below=.5ex of eve](k1){$\hspace{-.1ex}$\textcolor{black}{\small{$\mathbf{G}_{\mathsf{pub}}, t$}}};
	\node[above = 10.3ex of eve](inp){};
	\draw[<-,very thick,
	%decoration={
	%	zigzag,
	%	%				segment length=4,
	%	%				amplitude=.9,post=lineto,
	%	%				post length=2pt
	%},
	color=black](eve.north)--(inp);
	%	\node[rectangle, above right=2ex and 2ex of alice](t1){$\hspace{-.1ex}$\textcolor{black}{\small{ciphertex}}};			
	\node[rectangle, above right=-.4ex and 4ex  of alice](t2){$\hspace{-.3ex}$\textcolor{black}{\small{ciphertext:}}};	
	\node[rectangle, above right=-2.7ex and 1ex  of alice](t2){$\hspace{-.5ex}$\textcolor{black}{\small{$\mathbf{c} = \mathbf{m}\cdot \mathbf{G}_{\mathsf{pub}} + \mathbf{e}$}}};	
	\end{tikzpicture}
	\caption{Principle of the McEliece cryptosystem.}
	\label{fig:McEliece}
	\vspace{-1ex}
\end{figure}

The McEliece cryptosystem \cite{McEliece-PKC} and its dual variant, the Niederreiter cryptosystem \cite{Niederreiter-Cryptosystem}, provide general principles for code-based cryptography based on linear codes. The basic idea of the McEliece cryptosystem is illustrated in Fig.~\ref{fig:McEliece}. The private key consists of three matrices $\mathbf{S}$, $\mathbf{G}$, $\mathbf{P}$ where $\mathbf{G}$ is the generator matrix of a $t$-error correcting code. The private key is known only to the intended receiver ("Alice"). The public key is the {product} $\mathbf{G}_{\mathsf{pub}} = \mathbf{S} \cdot \mathbf{G} \cdot \mathbf{P}$ and $t$. Encryption (by "Bob") is done by calculating from a plaintext $\mathbf{m}$ the ciphertext $\mathbf{c} = \mathbf{m}\cdot \mathbf{G}_{\mathsf{pub}} + \mathbf{e}$, where $\mathbf{e}$ is a random vector of weight~$t$. The receiver ("Alice") can recover the plaintext because she knows the private key.
An eavesdropper ("Eve") should not be able to recover the matrix~$\mathbf{G}$ and thus has to decode an unknown code.

\section{The New Public-Key Cryptosystem based on Interleaved Goppa Codes}\label{sec:system}
\subsection{The System}\label{subsec:system}
Our proposed cryptosystem is an instantiation of the McEliece public-key cryptosystem using an $s$-interleaved \emph{non-binary} Goppa code $\mathcal{I}\Gamma(\mathcal{L},g,s)$ with parameters $[n, k\geq n-mr, d\geq r+1]_p$ and $p>2$.
 
This means that we encode $s$ messages of length $k$ into $s$ Goppa codewords of length $n$ and add a burst error of weight $\tIRS$ (see~\eqref{eq:tIRS}) to these interleaved codewords. 
Of course, we can see this as encrypting one message of length $sk$ into one ciphertext of length $sn$.
The decoding process that is needed for the decryption is then done in the IRS supercode $\IRSlong{n}{n-r}{s}$ which can correct a burst error of weight $\tIRS$.\\f

First, the key generation process is shown in the following. It is basically as for the McEliece cryptosystem based on Goppa codes, with the difference that an efficient decoder up to $\tIRS$ errors of the \emph{interleaved} Goppa code is now part of the private key.\\

%\textcolor{blue}{Careful with $k$, should be dimension of Goppa code, not RS code!}
\newpage
\textbf{Key Generation}:
\begin{itemize} 
	\item Choose parameters:
	\begin{itemize}
		\item Prime power $q=p^m$ with $p>2$;
		%	\item Irreducible monic polynomial $g(x)$ of degree $r$;
		\item Integers $m, n$ such that $ n \leq q$; %and $n \geq k \geq n - mr$;
		\item Locators $\mathcal{L} = (\alpha_1, \alpha_2, \dots, \alpha_n) \in \Fq^n$  ($n$ distinct elements of $\Fq$);
		\item Irreducible monic polynomial $g(x) \in \Fq[x]$ of degree $r$ such that $g(\alpha_i) \neq 0$, for all $i=1,\dots,n $;
		\item Integer $s$ and calculate $\tIRS = \lfloor \frac{sr}{s+1}  \rfloor$;
%		For a cryptosystem without decryption failure, choose $s=\tIRS=r-1$ (see Section~\ref{subsec:dec-guarant}).
	\end{itemize}
	\item Generate the following matrices:
	\begin{itemize}
		\item $\vec{S} \in \Fp^{k \times k}$: random non-singular matrix, called \emph{scrambler matrix};
		\item $\vec{G} \in \Fp^{k \times n}$: generator matrix of an $[n, k\geq n-mr, d\geq r+1]_p$ $\Gamma(\mathcal{L},g)$ code;
		\item $\vec{P}\in \Fp^{n \times n}$: random permutation matrix.
	\end{itemize}
	\item Compute $\Gpub = \vec{SGP} \in \Fp^{k \times n}$.
	\item Define the \emph{key pair}: 
	\begin{itemize}
		\item {Public key}: $(\Gpub,\tIRS, s)$;
		\item {Private key}: $(\vec{S}, \vec{P},\mathcal{D})$ where $\mathcal{D}$ is an efficient decoding algorithm for the $s$-interleaved Goppa code up to $\tIRS$ errors.\\
	\end{itemize}
\end{itemize}
Notice that we use $p>2$ since for $p=2$, the IRS supercode decoder cannot decode more errors than the standard Patterson decoder \cite{Patterson_GoppaDecoding} which can correct up to $r$ errors uniquely for \emph{binary} Goppa codes. \\

The encryption process works similar as in the McEliece cryptosystem, with the difference that we are now creating $s$ codewords of a Goppa code which are corrupted by a burst error of weight $\tIRS$.\\

\textbf{Encryption}:
\begin{itemize}
	\item {Input}: $s$ plaintexts $\vec{m}^{(i)} \in \Fp^k$, $i=1,\dots,s$;
	\item Generate full-rank random matrix $\vec{E}_{\tIRS} \in \Fp^{s\times \tIRS}$ without zero elements;
	\item Generate $\tIRS$ random positions $\mathcal{E} \subseteq \{1,\dots,n\}$ and denote by $\vec{e}^{(i)} \in \Fp^n$ the vector that contains the entries of the $i$-th row of $\vec{E}_{\tIRS}$ on the $\tIRS$ positions of $\mathcal{E}$ (ordered as in $\vec{E}_{\tIRS}$), for all $i=1,\dots,s$;
	\item Encryption: $\vec{c}^{(i)} = \vec{m}^{(i)} \Gpub + \vec{e}^{(i)} \in \Fp^n$, for all $i=1,\dots,s$;
	\item {Output}: $s$ ciphertexts $\vec{c}^{(i)} \in \Fp^n$, $i=1,\dots,s$.\\	
\end{itemize}
The reason to choose only non-zero elements for $\vec{E}$ is that having rows of smaller weight in $\vec{E}$ would facilitate an information set decoding attack on these rows and therefore decrease the overall security level (see also Section~\ref{sec:attacks}). Further, $\vec{E}_{\tIRS}$ is chosen to be a full-rank matrix as this decreases the failure probability of decoding/decryption significantly.\\

\textbf{Decryption}:
\begin{itemize}
	\item {Input}: $s$ ciphertexts $\vec{c}^{(i)} \in \Fp^n$, $i=1,\dots,s$;
	\item {Inverse Permutation}: $\vec{c}^{(i)}  \vec{P}^{-1} =  \mathbf{m}^{(i)} \vec{S} \vec{G} + \vec{e}^{(i)}  \vec{P}^{-1} $, for all $i=1,\dots,s$;
	\item {Decoding}: $(\vec{m}^{(1)} \vec{S},\dots,\vec{m}^{(s)} \vec{S} ) = \mathcal{D}(\mathbf{c}^{(1)}  \vec{P}^{-1},\dots,\mathbf{c}^{(s)}  \vec{P}^{-1}) $;
	\item $\vec{m}^{(i)} =  \vec{m}^{(i)}\vec{S}\vec{S}^{-1}$, for all $i=1,\dots,s$;
	\item {Output}: $s$ plaintexts $\vec{m}^{(i)} \in \Fp^k$, for all $i=1,\dots,s$.\\
\end{itemize}

\subsection{Decryption Guarantee}\label{subsec:dec-guarant}
The system from the previous section is shown as a general system, using \emph{any} integer $s$ as interleaving order. In the following, we will show that if we choose $s=r-1$ (which implies $\tIRS=r-1$), we can guarantee correct decryption (i.e., no decryption failures occur).
However, in Section~\ref{sec:attacks}, we show that for $s \geq \tIRS$, our system can be broken by the decoding algorithm from~\cite{MetznerKapturowski_1990}. 
Since we believe that failure-free decoding/decryption is interesting for the reader anyway, we show the details in the following.

As decoder in the decryption process, we can use any decoder of \cite{Bleichenbacher2003Decoding,Coppersmith2003Reconstructing,BrownMinderShokrollahi_ImprovedDecAGCodes-2005,Schmidt_CollaborativeDecoding_2009} which work for any $s$. In general, these decoders decode only with high probability. When choosing $s \geq \tIRS$, then, due to \cite[Thm.~2]{MetznerKapturowski_1990}, there is a unique decoding result and any interleaved decoder (e.g., the one from \cite{Schmidt_CollaborativeDecoding_2009}) \emph{guarantees} to return the unique decoding result.
In particular, if we want to have this guarantee and therefore no decryption failures, we can choose $s = \tIRS = r-1$.

%However, note that we will also show in the next section that this system with larger values of $s$ can be broken by the decoding algorithm from~\cite{MetznerKapturowski_1990}.

\begin{theorem}[Decoding/Decryption Guarantee]\label{thm:dec-guarantee}
	Consider an $s$-interleaved non-binary Goppa code $\mathcal{I}\Gamma(\mathcal{L},g,s)$ where each row is from a $\Gamma(\mathcal{L},g)$ code with parameters $[n, k\geq n-mr, d\geq r+1]_p$ and $p>2$. 
	Let $s = \tIRS = r-1$.
	
	For any $s \times \tIRS$ full-rank error matrix~$\vec{E}$, the syndrome-based interleaved decoder from~\cite{Schmidt_CollaborativeDecoding_2009} can always correct up to $\tIRS=r-1$ errors uniquely.
\end{theorem}
\begin{IEEEproof}
	The proof from \cite[Thm.~2]{MetznerKapturowski_1990} can be applied to our setting as follows.
	We prove that the syndrome matrix used in the decoding process of \cite{Schmidt_CollaborativeDecoding_2009} has full rank $\tIRS$ if $s = \tIRS$.
	
	Consider the following matrix:
	\begin{equation*}
	\vec{S}^\prime = \vec{H}^{\mathsf{RS}} \cdot \vec{E}^T,
	\end{equation*}
	where $\vec{E}$ is the $s \times n$ matrix from~\eqref{eq:matE} which is determined by the $s \times \tIRS$ full-rank matrix $\vec{E}_{\tIRS}$ and $\mathcal{E}$ in the encryption process; and $\vec{H}_{\mathsf{RS}}$ is a parity check matrix of the $\RS{n}{n-r}$ supercode from~\eqref{eq:RSsuper}.
	This can be rewritten by considering only the $\tIRS$ non-zero columns of $\vec{E}$ by
	\begin{equation*}
		\vec{S}^\prime = \vec{H}^{\mathsf{RS}}_{\tIRS} \cdot \vec{E}^T_{\tIRS},
	\end{equation*}
	where $\vec{E}_{\tIRS}$ contains the non-zero columns of $\vec{E}$ and $\vec{H}^{\mathsf{RS}}_{\tIRS}$ is the corresponding $(n-\kRS) \times \tIRS$ submatrix of $\vec{H}^{\mathsf{RS}}$.
	
	From \cite[Eq.~(17)]{Schmidt_CollaborativeDecoding_2009}, it follows that the IRS decoder always returns a unique decoding result (i.e., it does not fail) if $\vec{S}^\prime$ has rank $\tIRS$.
	This is true in our case since $\vec{E}_{\tIRS}$ has full-rank $\tIRS$ and $\vec{H}^{\mathsf{RS}}_{\tIRS}$ as well since it is a submatrix of a parity-check matrix. Thus, the decoder from \cite{Schmidt_CollaborativeDecoding_2009} will always correct up to $\tIRS=s$ errors.
\end{IEEEproof}

Notice that in the decryption process of our cryptosystem, the decoder is applied to $\vec{E}\vec{P}^{-1}$ which is a full rank matrix as both, $\vec{E}$ and $\vec{P}$ are full rank, and therefore Theorem~\ref{thm:dec-guarantee} holds for $\vec{E}\vec{P}^{-1}$ as well.

\section{Possible Attacks} \label{sec:attacks}
%\subsection{Structural Attacks}
%As the public key is the same as for the standard McEliece system based on Goppa codes~\cite{McEliece-PKC} which has remained secure against structural attacks for 40 years, it is quite unlikely that a structural attack on this proposed system exists.

\subsection{Metzner-Kapturowski Decoding Attack}\label{subsec:metzner-attack}
If $s \geq \tIRS$ and $\vec{E}_{\tIRS}$ has full rank, then the algorithm from \cite{MetznerKapturowski_1990}  decodes efficiently \emph{any interleaved code}, just by applying Gaussian elimination on the product of the parity-check matrix and the corrupted codeword.

This principle can therefore be applied as follows to break our system if $s \geq \tIRS$ and if $\vec{E}_{\tIRS}$ is a full-rank matrix:
\begin{itemize}
		\item Calculate $\Hpub$ from $\Gpub$ such that $\Hpub$ has full rank and $\Gpub\cdot \Hpub^T = \0$ (i.e., $\Hpub$ is a parity-check matrix of the public code generated by $\Gpub$);
	\item Consider the $s$ ciphertexts $\vec{c}^{(i)} \in \Fp^n$, $i=1,\dots,s$. They are codewords of the public code with generator matrix $\Gpub$ and parity-check matrix $\Hpub$, corrupted by a full-rank burst error.
	\item Decode all $\vec{c}^{(i)}$ with $\Hpub$ according to the algorithm from \cite{MetznerKapturowski_1990} which then provides the $s$ secret messages in cubic time.
\end{itemize}

This general decoding principle therefore provides an attack to our system if $s \geq \tIRS$ and if $\rk(\vec{E}_{\tIRS}) = \tIRS$. From a high-level point of view, decoding a "random" interleaved code is an easy problem for larger interleaving orders and if the burst errors has full rank.

Thus, for our parameter calculation in Section~\ref{sec:param}, we have to choose $s < \tIRS$.
Also, since it might be possible to search the whole solution space of the decoder from \cite{MetznerKapturowski_1990} if $s < \tIRS$, we should choose $s$ "not too close" to $\tIRS$. 
Straight-forward, searching the solution space would provide an attack of complexity $\mathcal{O} (n^3 p^{m(\tIRS-s)})$ which does not decrease the security level of the parameters that we suggest in Section~\ref{sec:param},
but the investigation if this can be done more efficiently is left for future work.

\subsection{Information Set Decoding and Ball Collision Attack}
In the classical \emph{information set decoding} (ISD) attack, the attacker tries to guess an error-free information set. 
The work factor of ISD clearly increases with the number of errors added in the encryption process.
Similar as it can be applied to the classical McEliece system based on Goppa codes, it can be applied to our interleaved system.

However, as we are adding a burst error (i.e., $s$ error vectors with non-zero elements at the same set of positions for each row) to the interleaved code, knowing an error-free information set of one ciphertext directly gives us error-free information sets of the other $s-1$ ciphertexts. Thus, we assume that the work factor (i.e., the average complexity) of ISD for our $s$-interleaved system is the same as for the classical McEliece system with \emph{one} ciphertext of length $n$ where $\tIRS$ arbitrary errors were added. 

For codes with $p\geq 3$, the work factor of the classical ISD attacks and its variants is larger than for $p=2$ as every operation needs to be carried out over $\Fp$. Therefore, to estimate the security level in bits (which is defined as the $\log_2$ of the work factor), the work factor for $p$-ary ISD is calculated by $\log_2(p)$ times the work factor for binary ISD for the same code parameters. This is just a lower bound and usually, the work factors for $p$-ary ISD can be much larger than for the binary case, see~\cite{Peters-ISDNon-binary}.

As best-known variant of ISD, we use the ball collision attack to calculate the work factors and security levels.
The work factor of the ball collision attack for \emph{binary Goppa codes} with parameters $[n,k,d]_2$ (when an error of weight $t$ is added) is:
\begin{equation}
\label{eq:bc}
\WBC(n,k,t) =  \min  \left \{ \frac{1}{2} \binom{n}{t} \binom{n-k}{t-\ell}^{-1} \binom{k}{\ell}^{-\frac{1}{2}} \ : \ 0 \leq \ell \leq \min \{t,k\}  \right \}.
\end{equation}

\subsection{Decoding-One-Out-of-Many Attack}\label{subsec:doom}
The decoding-one-out-of-many attack \cite{Sendrier-DecodingOneOutOfmany_2011} reduces the work factor $W$ of any attack if the attackers has access to $s$ ciphertexts and is interested in decoding only one of the messages. However, this attack provides a gain for the attacker on the work factor only if $ W \leq s^{3/2}$.
%with $T$ is the initial work factor or attack cost and $N$ is the number of available instances i.e. the interleaving order for interleaved Goppa code. 
%In our analysis, we therefore have to check if this condition holds. 
If this condition holds, e.g., the ISD work factor $\WISD$ would be reduced to  $\WISD^{2/3}$.
However, as we will see in the next section, this attack is not relevant for the parameters that we choose.

%\textcolor{blue}{[Is the last sentence also true if we only use small $s$????]}

%\textcolor{blue}{[TODO CHECK]}

\section{Comparison to Other Systems and Choice of Parameters}\label{sec:param}
To compare the security levels and key sizes of our system to known systems, we first compare the following three variants of Goppa codes used in the McEliece system:
\begin{enumerate}
	\item[A)] \textbf{$s$-Interleaved}: This refers to our suggested system, see Section~\ref{subsec:system}, where $s$ messages of length $k$ are encrypted into $s$ ciphertexts of length $n$ by a $\Gamma(\mathcal{L},g)$ Goppa code with parameters $[n,k,d\geq r+1]_p$ and a burst error of weight $\tIRS=\lfloor \frac{sr}{s+1}  \rfloor$ is added. 
	
	Table~\ref{tab:key} and Table~\ref{tab:seclevel} use interleaving orders with $s < \tIRS$  and therefore the system has a small decryption failure probability.
%	To give a decryption guarantee, we use $\tIRS = s = r-1$ in Table~\ref{tab:seclevel} (see Theorem~\ref{thm:dec-guarantee}).
	
	As explained in Section~\ref{sec:attacks}, the most efficient attack is the ball collision attack, which needs to be applied only to one ciphertext of length $n$.
	Thus, for a $p$-ary interleaved Goppa code, it has work factor (i.e., complexity) at least
	{$$\log_2(p)\WBC(n,k,\tIRS).$$}\\[-2ex]
	\item[B)] \textbf{$s$-Independent}: In this setting, also $s$ messages of length $k$ are encrypted into $s$ ciphertexts of length $n$ by an $[n,k,d\geq r+1]_p$ Goppa code. However, the errors are added on each ciphertext independently. The maximum number of errors that can be added to be able to decode each ciphertext independently uniquely is $\tU = \lfloor \frac{2r}{p} \rfloor$, see~\cite{Barreto_SquareFreeGoppa}. 
	To obtain guaranteed decoding in this scheme (i.e., failure probability zero), $\tU = \lfloor\frac{r}{2}\rfloor$ could be used. Notice that this decoding radius of \cite{Barreto_SquareFreeGoppa} is only achieved with a certain failure probability and only improves upon $\tU = \lfloor\frac{r}{2}\rfloor$ for $p=3$. 
	
	The work factor of the ball collision attack for the $s$ ciphertexts is therefore $s$ times the work factor for one ciphertext with $\tU$ errors, i.e., {$$s \log_2(p) \WBC(n,k,\tU).$$}
	This scheme can basically be seen as the classical McEliece cryptosystem (with a slightly better decoder if $\tU = \lfloor \frac{2r}{p} \rfloor$ is used).
	\\[-1ex]
	\item[C)] \textbf{Long Code}: Here, the message is encrypted to a single ciphertext of length $sn$ by using an $[n^* = sn, k^*, d^*\geq r^*+1]_p$ Goppa code which is constructed from a polynomial $g^*(x)$ of degree $r^*$. Decryption is done by using the unique decoder from~\cite{Barreto_SquareFreeGoppa} up to $\tU^* =  \lfloor \frac{2r^*}{p}  \rfloor$ errors. Also here, decryption is only done with high probability. The parameters $k^*$ and $d^*$ are chosen in our comparison either such that the key size is the same as in the $s$-interleaved case or such that the security level is the same. 
	
	The best known attack is the ball collision attack applied to the single ciphertext of length $n^* =sn$, i.e., it has work factor {$$\log_2(p)\WBC(sn,k^*,\tU^*).$$}\\[-2ex]
\end{enumerate}

Table \ref{tab:key} shows a comparison between the three systems A, B, C based on Goppa codes with (almost) equal key size where the key size in bits is
$K = \lceil \log_{2}(p)(n-k)k \rceil $ for Methods A and B, respectively $K^* = \lceil \log_{2}(p)(sn-k^*)k^* \rceil $ for Method~C. The security level in bits is the $\log_2$ of the previously mentioned work factors based on the ball collision attack. The dimension in the table is lower bounded by $k\geq n-rm$.

For Method C, there are two possibilities to choose $k^*$ as $K=K^*$ results in a quadratic equation in $k^*$, both resulting in the same security level but one of high code rate and one of low code rate.
For Method~A, in Table~\ref{tab:key} the general definition of the cryptosystem with small values for $s$ is used and therefore, the IRS decoder has a small failure probability (cf.~\cite{Schmidt_CollaborativeDecoding_2009}).
However, as both, Method~B and C are based on the decoding algorithm from~\cite{Barreto_SquareFreeGoppa}, they also have a decryption failure probability (which, according to our simulations, seems to be much larger than the one for Method~A based on \cite{Schmidt_CollaborativeDecoding_2009}).

We can see that for $s=3,7$, our proposed $s$-interleaved cryptosystem (Method~A) provides the largest security level.
The case $s=2$ and $p=3$ is quite special as the decoding radius $t$ coincides for methods A and B (and therefore adding burst errors actually does not make sense). However, in the interleaved case an attacker needs to decode only one out of the two ciphertexts and therefore the security level is lower than for Method B. 

In general, Method~A reaches a significantly larger security level for the same key size and therefore in the next table, only Method~A is compared to the classical McEliece system using list decoding.\\

%\newpage
\begin{table}[htb]
	\caption {Comparison between the different methods for codes with (almost) equal key size:\newline
Method A (\textbf{$s$-Interleaved}): $s \times [n , k, d]_{p}$ Goppa code with burst error of weight $t=\tIRS =  \lfloor \frac{sr}{s+1} \rfloor$ errors.\newline		 
Method B (\textbf{$s$-Independent}): $s$ codewords of an $[n , k, d]_{p}$ Goppa code with $s$ independent errors of weight $t = \lfloor \frac{2r}{p} \rfloor$.\newline 		 
		Method C (\textbf{Long Code}): a single codeword of an $[n^* = sn, k^*, d^*]_{p}$ Goppa code with an error of weight $t = \lfloor \frac{2r^*}{p} \rfloor$.
%		\textcolor{blue}{This table has a failure probability!}
		}
	\label{tab:key}
	\centering  
			\tabcolsep=0.23cm
	\begin{tabular}{ c ccccccccccc }
		\hline
		\hline\\[-1.5ex]
		Method & $s$ & $p$ & $r$ or $r^*$ & $t$ & $m$ & $k$ or $k^*$ $\geq$ & $n$ or $n^*$&  Security level [bits] & Code rate $\geq$ & Key size [bits] \\
		\hline
		\hline\\[-1.5ex]
		\textbf{A (Interleaved)}& 2 & 3 & 54 & 36 & 7 & 1809 & 2187 & {\textbf{84}} & \textbf{0.82} & \textbf{1083801} \\
		B & 2 & 3 & 54 & 36 & 7 & 1809 & 2187 &  86 & 0.82 &  1083801\\
		C & 2 & 3 & 21 & 14 & 8 & 4211 & 4374 &  58 & 0.96 & 1087908 \\
		C & 2 & 3 & 527 & 351 & 8 & 163 & 4374 & 58 & 0.03 & 1087908 \\  
		\hline\\[-1.5ex]
		\textbf{A (Interleaved)}& 3 & 3 & 54 & 40 & 7 & 1809 & 2187 & {\textbf{93}} & \textbf{0.82} & \textbf{1083801} \\
		B & 3 & 3 & 54 & 36 & 7 & 1809 & 2187 & 86 & 0.82 & 1083801 \\
		C & 3 & 3 & 14 & 9 & 8 & 6455 & 6561 & 45 & 0.98 & 1084479 \\
		C & 3 & 3 & 807 & 538 & 9 & 106 & 6561 & 45 & 0.01 & 1084479 \\  
		\hline\\[-1.5ex]
		\textbf{A (Interleaved)}& 7 & 3 & 54 & 47 & 10 & 58509 & 59049 &  {\textbf{256}} & \textbf{0.99} & \textbf{50076669} \\
		B & 7 & 3 & 54 & 36 & 10 & 58509 & 59049 & 200 & 0.99 & 50551776 \\
		C & 7 & 3 & 7  & 7  & 12 & 413267 & 59049 & 37  & 0.99 & 50435856 \\
		C & 7 & 3 & 34439 & 34439 & 12 & 76 & 59049 & 6 & 0.001 & 50435856 \\
%		\textcolor{blue}{Add C}\\
		\hline
	\end{tabular}
\end{table}

\newpage
\vspace{4ex}
%Table \ref{tab:seclevel} shows a comparison between the methods for different Goppa codes with (almost) equal security level. 

%Our $s$-interleaved scheme reduces the key size significantly compared to the $s$-independent (Method~B) scheme and the long code (Method~C). 

In Table \ref{tab:seclevel}, we compare our scheme with using a binary Goppa code and list decoding as presented in \cite{Barbier_2011} which is the most efficient instantiation of the McEliece system based on \emph{standard} (i.e., square-free or irreducible) Goppa codes. Compared to the McEliece system based on unique decoding, list decoding as in \cite{Barbier_2011} reduces the key size by at most 5\% for the same security level.
As we have seen in Table~\ref{tab:key}, our $s$-interleaved scheme reduces the key size significantly compared to the $s$-independent (Method~B) scheme and the long code (Method~C). Those two methods and unique decoding of Goppa codes (i.e., the "classical" McEliece scheme) are therefore not listed in Table~\ref{tab:seclevel}.

%However, QC-MDPC codes do not provide a decryption guarantee as we do (since in Table~\ref{tab:seclevel}, we chose $s = \tIRS = r-1$, see also Theorem~\ref{thm:dec-guarantee}) and therefore have a non-negligible failure probability which enables reaction-based attacks such as \cite{GuoJohanssonStankovski-KeyRecoveryMDPC}.
Note that our decryption algorithm (Method~A) has a small failure probability.
As previously mentioned, Methods~B and~C have a non-negligible failure probability as well since they are based on the decoding algorithm from~\cite{Barreto_SquareFreeGoppa}. According to our simulations, the failure probability of interleaved decoding seems to be much smaller than the one from~\cite{Barreto_SquareFreeGoppa}.
%This failure rate was conjectured in~\cite{Barreto_SquareFreeGoppa} to be roughly $\frac{1}{p^{m(r+1-t)}}$ which is also shown in Table~\ref{tab:seclevel}.
The list decoding method from~\cite{Barbier_2011} might return a list of possible codewords which can be seen similar to a failure probability. Note that their list might be reducible to one codeword by using a CCA2-secure variant, see~\cite{Barbier_2011}, but this requires additional effort (rate reduction) and is not considered here.

The decoding-one-out-of-many attack mentioned in Section~\ref{subsec:doom} does not play a role in both tables as the work factor for all methods is always larger than $s^{3/2}$.

Note that there are other schemes, e.g., quasi-cyclic moderate-density-parity-check (QC-MDPC) codes as in \cite{Misoczki_2013}, where our key size is larger for the same security level. However, as QC-MDPC codes have not resisted structural attacks for 40 years as Goppa codes, we believe it is fair to mainly compare our results to \cite{Barbier_2011}.

Our Method~A reduces the key size significantly compared to using list decoding in the classical McEliece system. In particular, for security level 128 bit, the bottom-most instantiation is remarkable as the code rate and security level are the same as in \cite{Barbier_2011}, the total ciphertext length $sn = 2662$ is smaller than the one of \cite{Barbier_2011} (3262), and the key size is reduced by 42\%.
Similary, for security level 256 bit, the bottom-most row shows almost the same code rate, same security level, $sn = 8788$ is only slightly larger than $n=7008$, but the key size is reduced by 59\%. Even larger reductions of the key size are possible when we allow $sn$ of Method~A to be larger than $n$ of \cite{Barbier_2011}, e.g., for security level 128 bit, the third instantiation of our method reduces the key size compared to \cite{Barbier_2011} by 64\%.

\begin{table}[htb!]
	\caption {Comparison between different schemes leading to (almost) equal security level:\newline
Method A (\textbf{$s$-Interleaved}): $s \times [n , k, d]_{p}$ Goppa code with burst error of weight $t=\tIRS = \lfloor \frac{sr}{s+1} \rfloor$ errors.\newline		 
\textbf{List decoding} \cite{Barbier_2011}: one codeword of an $[n , k, d \geq 2r+1]_{2}$ binary Goppa code where $t$ is the binary Johnson radius: $t = \frac{n}{2}(1-\sqrt{1-\frac{4r+2}{n}})$.\newline 		 
%Method C (\textbf{Long Code}): a single codeword of an $[n^* = sn, k^*, d^*]_{p}$ Goppa code with an error of weight $t = \lfloor \frac{2r^*}{p} \rfloor$.\newline
%		Method E: big code $[s \times n, s \times k, d']_{p} \quad \textrm{with} \quad  t' = \left\lfloor \dfrac{2r'}{p} \right\rfloor$
%				\begin{itemize}		
%					\item 
%				\textcolor{blue}{WHAT DOES THIS MEAN??	[*] not exactly the ISD-attack (to a factor of the co-dimension $n-k$), however used to define the security level of the code.
%					\item
%					 [**] key size = length - co-dimension = dimension}
%					\item
%					 [***] $P_{1s}$: Pr( rank(E) != s) with simulation
%					\item
%					 $X_{n \times s}$: lower bound on the number of full-rank full-weight matrices
%				\end{itemize}		
		}  
	\label{tab:seclevel}
%	\begin{threeparttable}
		\tabcolsep=0.22cm 
		\centering  
		\begin{tabular}{ c cccccccccccccc}
			\hline
			\hline\\[-1.5ex]
			Method & $s$ & $p$ & $r$ & $t$ & $m$ & $k \geq n-rm$ & Length $n$ &  Security level [bits] & Code rate $\geq$ & Key size [bits]\\%& Failure probability \\%& $X_{n \times s}$ & $\frac{3}{2} \log_2{(s)}$ & $P_{1s}$\\
			\hline   \\[-1.5ex]
			List decoding & 1 & 2 & 40 & 41 & 11 & 1436 & 1876 & 80 & 0.77 & 631 840\\
			\textbf{A (Interleaved)} & 21 & 4 & 42 & 40 & 5 & 814 & 1024 & \textbf{80} & \textbf{0.79} & \textbf{341 880}\\
			\textbf{A (Interleaved)} & 6 & 4 & 54 & 46 & 5 & 754 & 1024 & \textbf{80} & \textbf{0.74} & \textbf{407 160}\\			
			
%			\hline   \\[-1.5ex]
%			List decoding & 1 & 2 & 58 & 59 & 12 & 2196 & 2868 & 112 &  0.77 & 1475 712 \\ 
%			\textbf{A (Interleaved)}& 44 & 3 & 49 & 47 & 7 & 1844 &2187& \textbf{112} & \textbf{0.84} & \textbf{1002 476} \\
%			B & 14 & 3 & 44 & 29 & 8 & 6209 & 6561 & 112 & 0.94 & 3464 044 \\ %$\approx 8.4 \cdot 10^{-62}$ \\ 
%			C & 2 & 3 & 19 & 12 & 14 & 6209 & 6561 & 116 & 0.94 & 1304751 730 \\
			\hline\\[-1.5ex]
%			QC-MDPC  & 1 & 2 & - & 142 & - & 9857 & 19714 & 128 & 0.5 & $9857$\\ %& $< 10^{-7}$, but $>0$ \\
			List decoding& 1 & 2 & 65 & 66 & 12 & 2482 & 3262 & 128 &  0.76 & 1935 960\\ 
			\textbf{A (Interleaved)}& 31 & 3& 64 & 62 & 7 &1739 &2187 & \textbf{128} & \textbf{0.76} &  \textbf{1234 799}\\%($P'$) &$ > 2^{464}$ & 9 & 0.4338 \\
			\textbf{A (Interleaved)}& 7 & 3& 84 & 73 & 7 & 1599 &2187 & \textbf{128} & \textbf{0.73} &  \textbf{1490 200}\\
			\textbf{A (Interleaved)}& 20 & 11 & 58 & 55 & 3 & 1157 & 1331 & \textbf{129} & \textbf{0.87} & \textbf{696 445}\\
			\textbf{A (Interleaved)}& 2 & 11 & 107 & 71 & 3 & 1010 & 1331 & \textbf{127} & \textbf{0.76} & \textbf{1121 582}\\
%			\hline\\[-1.5ex]
%			List decoding & 1 & 2 & - & 96 & 13 & 4021 & 5269 & 192 &  0.75 & 5018208 & list \\ 
%			\textbf{A (Interleaved)}& 60 & 3 & 61 & 60 & 8 & 6073 & 6561 & \textbf{193} & \textbf{0.92} & \textbf{4697233} & \textbf{0} \\%($P'$) & $ > 2^{548}$ & 8.86 & 0.4562 \\ 
			\hline\\[-1.5ex]
%			QC-MDPC & 1 & 2 & 274 & 264 & - & 32771 & 65542 & 256 & 0.5 & 32771 \\%& $< 10^{-7}$, but $>0$\\
			List decoding & 1 & 2 & 130 & 133 & 13 & 5318 & 7008 & 257 &  0.76 & 8987 420 \\% & list \\ 
			\textbf{A (Interleaved)} &  10 & 5 & 167 & 151 & 5 & 2290 & 3125 & \textbf{256} & \textbf{0.73} & \textbf{4439 874} \\
			\textbf{A (Interleaved)} &  6 &  5 & 206 & 176 & 5 &2059 & 3125 & \textbf{256} & \textbf{0.67} & \textbf{5010 372}\\
			\textbf{A (Interleaved)} & 65 & 13 & 131 & 129 & 3 & 1804 & 2197 & \textbf{257} & \textbf{0.82} & \textbf{2623 508}\\
			\textbf{A (Interleaved)} & 4 & 13 & 207 & 165 & 3 & 1576 & 2197 & \textbf{257} & \textbf{0.72} & \textbf{3621 605}\\
			\hline
		\end{tabular}
%		\begin{tablenotes}\footnotesize
%			
%			\item [*] not exactly the ISD-attack (to a factor of the co-dimension $n-k$), however used to define the security level of the code.
%			\item [**] key size = length - co-dimension = dimension
%			\item [***] $P_{1s}$: Pr( rank(E) != s) with simulation
%			\item $X_{n \times s}$: lower bound on the number of full-rank full-weight matrices
%		\end{tablenotes}
%	\end{threeparttable}
\end{table}

\newpage
\section{Conclusion and Outlook}\label{sec:concl}
In this paper, we have presented a public-key code-based cryptosystem based on interleaved Goppa codes. 
Compared to the classical McEliece system using list decoding, our scheme can reduce the key size by over 50\% for the same security level while keeping the code structure which has remained resilient against structural attacks for many years.\\

The following tasks are left for future work:
\begin{itemize}
\item Explicit analysis of the failure probability of our approach. This includes the derivation of an explicit upper bound as in \cite{Schmidt_CollaborativeDecoding_2009} for decoding interleaved subfield subcodes (e.g., Goppa codes) as well the simulation of the failure probability for the parameters suggested in Tables~\ref{tab:key} and \ref{tab:seclevel};
\item Can the decoding algorithm from \cite{MetznerKapturowski_1990} be used to obtain an efficient attack by searching the solution space? Straight-forward, the complexity is too large to reduce the security level (see Section~\ref{subsec:metzner-attack});
\item Interleaved decoding of Goppa codes not only to the interleaved decoding radius of the IRS code, but to a larger radius.
\end{itemize}

%Outlook: failure probability comparison, attack based on Metzner (exponential search) if $s <t$ but $s$ close to $t$, decoding interleaved Goppa codes not only in IRS supercode

\section*{Acknowledgment}
The authors would like to thank Lukas Holzbaur, Sven Puchinger, and Vladimir Sidorenko for the valuable discussions.

\bibliographystyle{IEEEtranS}
\bibliography{main}

\end{document}

%% file: defs.tex
 \newtheorem{theorem}{Theorem}
 \newtheorem{definition}{Definition}

\DeclareMathOperator{\wt}{wt}

\DeclareMathOperator{\rk}{rk}

\newcommand{\Fq}{\mathbb F_{q}}
\newcommand{\Fp}{\mathbb F_{p}}

\newcommand{\RS}[2]{\ensuremath{\mathcal{RS}(#1,#2)}}
\newcommand{\IRS}[1]{\ensuremath{\mathcal{IRS}(n,\kRS,#1)}}
\newcommand{\IRSlong}[3]{\ensuremath{\mathcal{IRS}(#1,#2,#3)}}

\newcommand{\E}{\mathbf E}

\newcommand{\0}{\mathbf 0}
\newcommand{\printalgo}[1]

\renewcommand{\vec}[1]{\mathbf{#1}}
\newcommand{\kRS}{k_{\mathsf{RS}}}
\newcommand{\tIRS}{t_{\mathsf{IRS}}}
\newcommand{\tU}{t_{\mathsf{u}}}
\newcommand{\Gpub}{\mathbf{G}_{\mathsf{pub}}}
\newcommand{\Hpub}{\mathbf{H}_{\mathsf{pub}}}
\newcommand{\WBC}{W_{\mathsf{BC}}}
\newcommand{\WISD}{W_{\mathsf{ISD}}}